# Measuring Bullet Velocity with a PC Soundcard

Michael Courtney and Brian Edwards

Abstract:
This article describes a simple method for using a PC soundcard to accurately measure bullet velocity.  The method involves placing the microphone within a foot of the muzzle and firing at a steel target between 50 and 100 yards away.  The time of flight for the bullet is simply the recorded time between muzzle blast and sound of the bullet hitting the target minus the time it takes the sound to return from the target to the microphone.  The average bullet velocity is simply the distance from the muzzle to the target divided by the time of flight of the bullet.  This method can also be applied to measurement of paintball velocities.

**Introduction**
Physics has a long and important relationship with the measurement of bullet velocity. The earliest accurate measurements of bullet velocity relied on the ballistic pendulum. The ballistic pendulum remains an important laboratory tool for demonstrating conservation of momentum and energy, as well as providing a method for measuring the velocity generated by potato cannons.[1]

As firearms design matured, higher velocities made capture of a bullet in a pendulum more difficult.  Longer range applications from higher velocity and increased accuracy also mandated more accurate velocity measurements in order to accurately predict long range trajectories.  This led to the development of the optical chronograph (commonly known simply as the chronograph in modern ballistics) which measures bullet velocity by optically detecting the bullet passing over fixed points and using the time interval to compute the velocity.

As paintball has become increasingly popular, most paintball facilities require that paintball markers (the devices which shoot paintballs are called "markers" rather than "guns") be adjusted to keep the projectile velocity under 250 feet per second.   This ensures the safety of participants, since injury becomes more likely as velocities exceed 250 feet per second.  Optical chronographs have long been used in this application, but recently less expensive paintball chronographs have become available that employ the Doppler effect.

Measuring the velocity of a bullet or paintball with an optical chronograph is a relatively simple process and the price of these instruments has become more affordable in the last several years.  However, shooters who use their chronographs regularly and under a variety of conditions have probably noticed some drawbacks.  This technique requires putting the instrument in front of the gun and shooting through skyscreens, which contain a photodiode, a lens, and light diffuser.  Some models require the electronics to also be placed in front of the gun.

Consequently, replacing skyscreens and electronics can become a recurring operational expense, especially if one's aim is poor or if one regularly chronographs shotgun loads or saboted muzzleloader projectiles.  Shotgun wads and plastic sabots can quickly separate from the projectile(s) and damage the equipment.  An important rule of firearms safety is never allow the muzzle to point at anything you are not willing to destroy.  Many chronographs end up vivid demonstrations of this important safety rule.

Optical chronographs can also be inconsistent in some lighting conditions such as late afternoon, early morning, or the covered shooting area of a range.  Cables and sky screens can be an irritation to set-up.  Fiddling around with a chronograph in front of the firing line is an unwelcome activity at many range facilities.

This article describes a simple method for using a PC soundcard to accurately measure bullet or paintball velocity.  Soundcards are inexpensive and readily available, and the most computers today include a soundcard.  The biggest advantage of the method is that no expensive equipment needs to be placed in front of the gun.  Non-marking (paintless) projectiles are available to measure the muzzle velocity of paintball markers with less mess.

**The Soundcard Method**
The soundcard method involves placing the microphone within a foot of the muzzle, preferably a bit to the side and slightly behind the muzzle.  A steel target that makes a loud noise when hit is placed at a carefully measured distance between 50 and 100 yards away.  Targets that make a resonating sound like a gong work better than targets which make more "clink" type of sounds.  For paintball markers, much shorter distances (15-30 feet) should be used.

The soundcard should be operated with software (such as Audacity[2]) that allows the user to view the sound waveform and accurately determine the time difference between the beginning of the muzzle blast and the beginning of the sound of the bullet hitting the steel target.  One can also use a Vernier LabPro with a microphone.  Under some conditions, especially if the range is covered and the target is close, reverberation of the muzzle blast will be loud enough to mask the sound of the bullet hitting the target.  If this occurs, the easiest solution is increasing the distance to the target to give the muzzle blast more time to dissipate before the sound of the bullet hitting the target arrives back at the microphone.

The time of flight of the bullet ($t_{bullet}$) is determined by subtracting the time it takes sound to travel from the target back to the microphone ($t_{sound}$) from the time recorded between the muzzle blast and the sound of the bullet hitting the target ($t_{total}$).  Expressed as a formula,

$t_{bullet} = t_{total} - t_{sound}$.

The time it takes the sound to travel back from the steel target to the microphone is simply the distance (*d*) divided by the velocity of sound ($V_{sound}$),

$t_{sound} = d / V_{sound}$.

*Table 1: The velocity of sound for temperatures from 0 to 100 ° F.*

| Temperature (° F) | $V_{sound}$ (ft/s) | Temperature (° F) | $V_{sound}$ (ft/s) |
|---|---|---|---|
| 0 | 1049.7 | 50 | 1105.4 |
| 2 | 1052.0 | 52 | 1107.6 |
| 4 | 1054.3 | 54 | 1109.7 |
| 6 | 1056.6 | 56 | 1111.9 |
| 8 | 1058.8 | 58 | 1114.0 |
| 10 | 1061.1 | 60 | 1116.2 |
| 12 | 1063.4 | 62 | 1118.3 |
| 14 | 1065.6 | 64 | 1120.5 |
| 16 | 1067.9 | 66 | 1122.6 |
| 18 | 1070.1 | 68 | 1124.7 |
| 20 | 1072.3 | 70 | 1126.9 |
| 22 | 1074.6 | 72 | 1129.0 |
| 24 | 1076.8 | 74 | 1131.1 |
| 26 | 1079.0 | 76 | 1133.2 |
| 28 | 1081.3 | 78 | 1135.4 |
| 30 | 1083.5 | 80 | 1137.5 |
| 32 | 1085.7 | 82 | 1139.6 |
| 34 | 1087.9 | 84 | 1141.7 |
| 36 | 1090.1 | 86 | 1143.8 |
| 38 | 1092.3 | 88 | 1145.9 |
| 40 | 1094.5 | 90 | 1148.0 |
| 42 | 1096.7 | 92 | 1150.1 |
| 44 | 1098.9 | 94 | 1152.1 |
| 46 | 1101.0 | 96 | 1154.2 |
| 48 | 1103.2 | 98 | 1156.3 |
| 50 | 1105.4 | 100 | 1158.4 |

The velocity of sound is roughly 1100 feet per second (*ft/s*), but it varies with temperature as shown in Table 1. In addition to having an accurate distance measurement (tape measure rather than range finder), one needs the ambient temperature within 2 degrees to have an accurate $V_{sound}$.

Once the time of flight of the bullet is determined, the average velocity of the bullet is $V_{bullet} = d / t_{bullet}$.

**Results and Analysis**
The first load tested is a mild .223 Remington load. The resulting sound waveform is shown in Figure 1. Our software is set to begin recording the waveform when a loud noise is detected, so the beginning of the muzzle blast is at t = 0. As shown in Figure 1, the sound level saturates the equipment for the first 0.04 seconds, and there are some prominent peaks intermittently through the signal that correspond to echoes. There is a large peak at *t = 0.1883s* that records the bullet hitting the steel target.

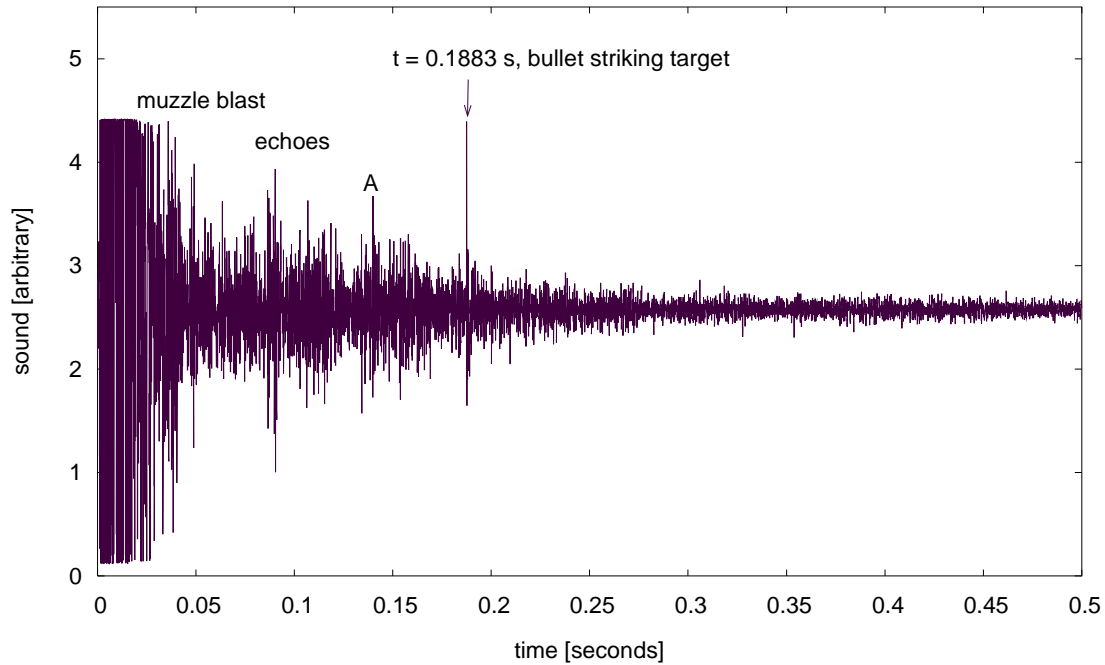

Figure 1: .223 Remington Sound Waveform

We shot a metal target located 150 feet away. The speed of sound at *82° F* is *1139.6 ft/sec*. Consequently, the time for the sound to return from the target to the microphone is $t_{sound} = 150\ ft/(1139.6\ ft/s) = 0.1316\ s$.

Analysis of the sound waveform shown in Figure 1 (zooming in is required) shows the time between the muzzle blast and sound of the bullet hitting the target is $t_{total} = 0.1875\ s$. Subtracting the time it takes the sound to return from the target to the microphone gives

$t_{bullet} = t_{total} - t_{sound} = 0.1875\ s - 0.1316\ s = 0.0559\ s.$

The average velocity of the bullet over the distance of 50 yards is then

$V_{bullet} = d / t_{bullet} = 150\ ft / (0.0559\ s) = 2684\ ft/s.$

For comparison, the velocity measured on an Oehler Model 35 optical chronograph placed 6 feet in front of the muzzle is *2763 ft/s*. At first glance, these measurements do not seem to agree, but it is important to note that our acoustic method measures the average velocity between the bullet and the target. This average velocity is slightly slower than the muzzle velocity because the bullet slows down in flight as a result of air resistance.

There are a number of public access ballistics programs that one can use for computing time of flight and trajectory for a given muzzle velocity and ballistic coefficient. The ballistic coefficient is the specification describing how air resistance slows down a bullet in flight. The ballistic coefficient is a dimensionless quantity that roughly corresponds to the fraction of 1000 yards that a bullet must travel to lose 50% of its initial kinetic energy (or 29.3% of its initial velocity). For example, a bullet with a ballistic coefficient of 0.4 has lost 50% of its kinetic energy at a range of 400 yards. Performing a Google search for "ballistic calculator" will provide several options. We have found the JBM ballistic calculator reliable and easy to use.[3]

Using the manufacturer's published ballistic coefficient of *0.200* for the .223 bullet, the JBM ballistics calculator predicts a time of flight of *0.056 s* for 50 yards, thus an average velocity *2679 ft/s* for a *50 yard* flight. Consequently, our acoustic measurement of bullet velocity is in excellent agreement with the optical chronograph, with a relative error of 0.2%.

Looking again at Figure 1, one might be tempted to misinterpret one of the earlier echo peaks as corresponding to the sound of the bullet hitting the plate, but more careful analysis rules out this possibility. For example, consider the peak labeled "A" with a magnitude close to *3.7* located at a time of *0.1400 s*. Since sound takes *0.1316 s* to return from the target, this peak would correspond to a bullet time of flight of *0.0084 s* suggesting an impossible velocity over *17,000 ft/s*.

We also tested the acoustic method of measuring bullet velocity using a 38 grain hollow point load in the 22 Long Rifle cartridge (22LR). A sound waveform is shown in Figure 2. The sound level saturates the equipment for the first *0.025* seconds, and there are some prominent peaks between *0.06 s* and *0.12 s* corresponding to echoes. As explained above, it is not possible to interpret these early peaks as the sound of the bullet hitting the target because it takes sound over *0.13 s* to return from the target.

There is a large peak with a magnitude of *3.5* at $t = 0.2594\ s$. This is the peak that records the bullet hitting the steel target.

The speed of sound at *76° F* is *1133.2 ft/s*. The time for the sound to return from the target to the microphone is $t_{sound} = 150\ ft/(1113.2\ ft/s) = 0.1324\ s$.

The total time between the muzzle blast and sound of the bullet hitting the target is $t_{total} = 0.2594\ s$. Subtracting the time it takes the sound to return from the target to the microphone gives

$t_{bullet} = t_{total} - t_{sound} = 0.2594s - 0.1324s = 0.1270s.$

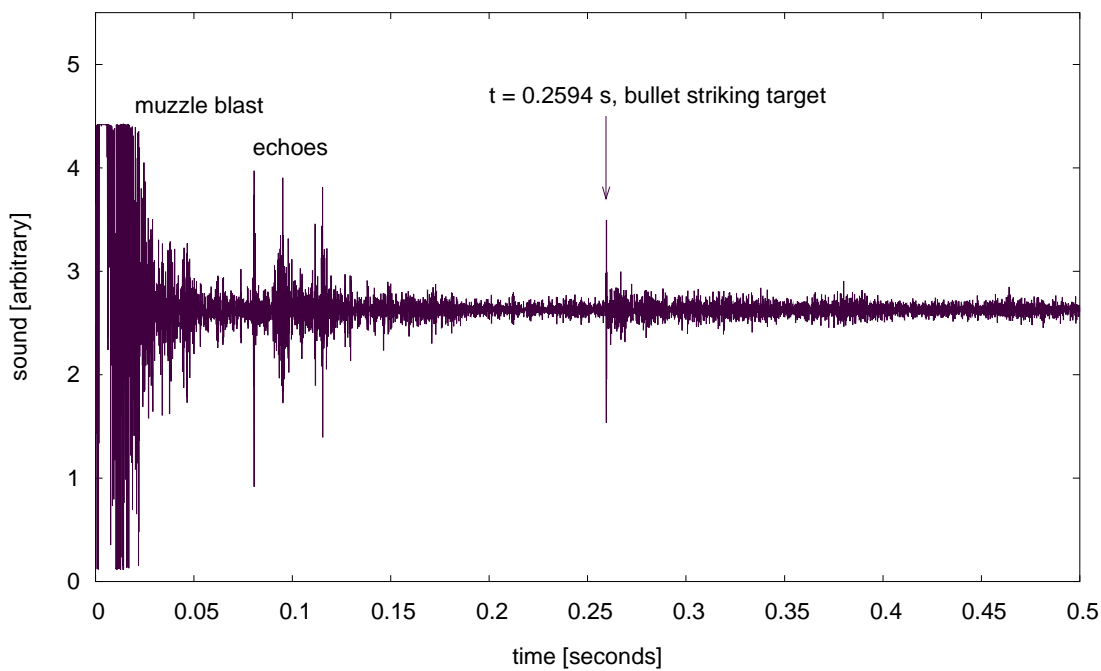

Figure 2: 22 Long Rifle Sound Waveform

The average velocity of the bullet over the distance of 50 yards is then

$V_{bullet} = d / t_{bullet} = 150 ft/ (0.1270s) = 1181 ft/s.$

The optical chronograph velocity for this shot is *1270 ft/s*. As expected, the average velocity from the acoustic method is slightly slower than the chronographed muzzle velocity because the bullet slows down in flight. Using a ballistic coefficient of 0.124, we compute the time of flight for the measured chronograph velocity to predict a time of flight of *0.126 s* for *50 yards,* thus an average velocity *1190 ft/s*. The acoustic measurement of bullet velocity is in excellent agreement with a chronograph with a relative error of 0.8%.

Since this technique is partly motivated by the propensity for muzzleloaders to be tough on chronographs (sabots and corrosive residue), we also tested the acoustic method on a muzzleloader. Our test load was 40 grains of Hodgdon Triple Seven powder in a 45 caliber muzzleloader behind a saboted .357 caliber 180 grain bullet.

Our test load is relatively quiet. Many muzzleloader loads have a muzzle blast of significantly longer duration that requires placing the target at 100 yards to prevent the muzzle blast from obscuring the sound of the bullet hitting the target in the sound waveform. We were able to test this milder load at 50 yards.

As shown in Figure 3, the sound level saturates the equipment for the first *0.05 seconds*, and there are some prominent peaks in the *0.06 to 0.17 second* range that correspond to echoes. We cannot interpret these peaks as the sound of the bullet hitting the target because they occur either before sound could even travel *50 yards*, or they correspond to velocities over *3000 ft/s*. The large peak with magnitude of *3.8* at $t = 0.2487$ s corresponds to the bullet hitting the steel target.

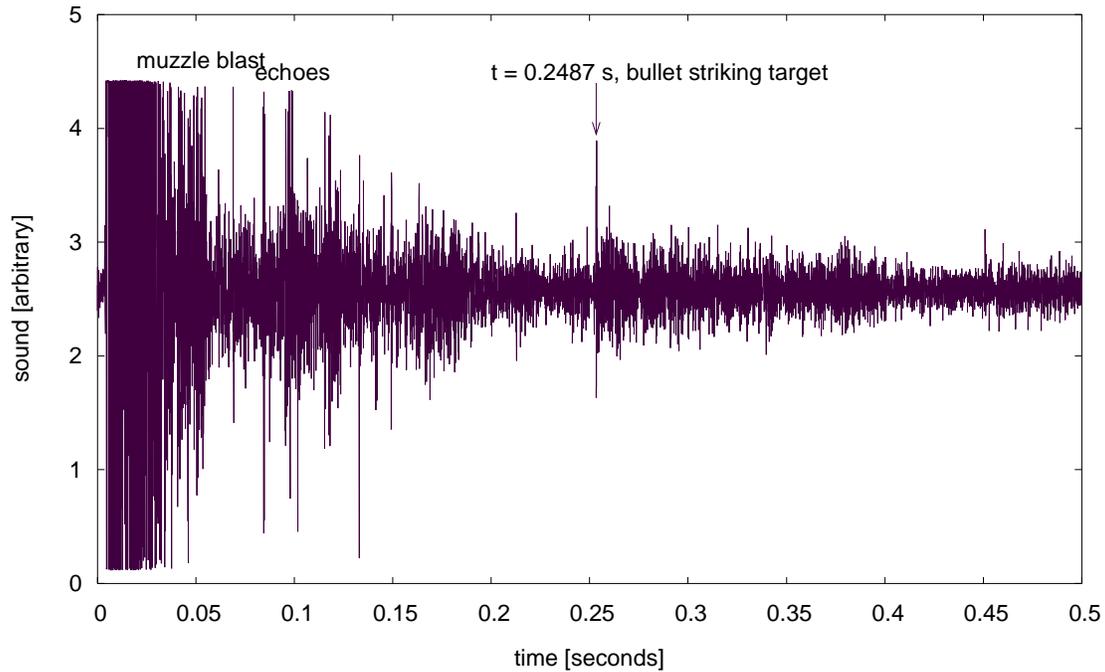

Figure 3: 45 Caliber Muzzleloader Sound Waveform

From Table 1, the speed of sound under the ambient conditions of *78° F* is *1135.4 ft/s*. The time for the sound to return from the target to the microphone is $t_{sound} = 150$ *ft/ (1135.4 ft/s) = 0.1321 s*.

The total time between the muzzle blast and sound of the bullet hitting the target is $t_{total} =$ *0.2487 s*. Subtracting the time it takes the sound to return from the target to the microphone gives

$t_{bullet} = t_{total} - t_{sound}$ = *0.2487 s – 0.1321 s = 0.1166 s.*

The average velocity of the bullet over the distance of 50 yards is then

$V_{bullet} = d / t_{bullet}$ = *150 ft/ (0.1166 s) = 1286 ft/s.*

For comparison, the velocity measured on the optical chronograph is *1328 ft/s*. Once again, this average velocity is slightly slower than the muzzle velocity because the bullet slows down in flight as a result of air resistance. In this case, detailed comparison with the chronograph is not possible because the ballistic coefficient of the projectile changes mid-flight when the plastic sabot separates from the bullet.

**Discussion and Conclusions**
One might consider using this acoustic time of flight measurement technique along with an optical chronograph to measure an unknown ballistic coefficient. Many ballistic software programs will compute the ballistic coefficient given the muzzle velocity from a chronograph and time of flight from the soundcard. If the software program lacks this feature, one might guess different ballistic coefficients until the software gives the measured time of flight. We should caution those considering this approach that very small errors in the time of flight determination yield relatively large errors in the determination of ballistic coefficients.

Even though the acoustic technique described above can measure the time of flight and bullet velocity reliably with an accuracy better than 1%, attempts to determine an unknown ballistic coefficient by this method can be off by as much as 10-20% due to uncertainties introduced by errors in the distance measurement and speed of sound determination.

Accurately determining the time of flight well enough to reasonably estimate the ballistic coefficient requires both measuring the distance to the target within a couple of inches and placing a second microphone within two inches of the target and recording the time difference between the muzzle blast on the near microphone and the sound of impact on the far microphone. The safest place for the far microphone is behind an impenetrable steel target.

We believe that the soundcard method of measuring bullet velocity has potential commercial software application in a program that uses a soundcard and automates the processes described above. However, reliable commercial application will require discerning the sound of the bullet hitting the target even in cases where the muzzle blast and other sounds prevent the clear discrimination of a peak when the bullet hits the target. This can be accomplished with standard techniques of Fourier analysis where the program learns the unique spectral signature of the target and detects the time at which this spectral signature occurs in the sound waveform. The Audacity software program has a sound spectrogram feature that facilitates this and can be used to demonstrate the benefits of a resonant target.

In conclusion, we have presented a useful method for measuring bullet velocity with a PC soundcard or Vernier LabPro. This method provides reasonable accuracy (*1%*) but requires some care. Admittedly our method of measuring bullet velocity with a soundcard has some drawbacks. Loud noise can interfere with the measurement. The technique requires some computations, but these are relatively simple and are not hard to set up in a spreadsheet program. On the positive side, the PC soundcard method of measuring bullet velocity is a great example of Physics and technology working together to accomplish a technically challenging goal without putting specialized equipment in front of the barrel of a gun.

*About the Authors:*

Michael Courtney has a PhD in experimental Physics from the Massachusetts Institute of Technology and has authored and co-authored numerous papers in the peer-reviewed journals.  Michael worked for seven years as a microwave test and measurement engineer for high-tech companies before becoming an Assistant Professor of Physics at Lorain County Community College.  He is an NRA certified instructor for Basic Pistol and Personal Protection in the Home as well as a volunteer hunter education instructor for the Ohio Division of Wildlife.  Division of Science and Math, Lorain County Community College, Elyria, OH 44035; Michael_Courtney@alum.mit.edu

Brian Edwards is a member of the NRA who is an avid hunter and firearms enthusiast. He is an exceptional Physics student who hopes to major in Physics or Engineering. Division of Science and Math, Lorain County Community College, Elyria, OH 44035.